# The Association Between SOC and Land Prices Considering Spatial Heterogeneity Based on Finite Mixture Modeling

유한혼합모델링을 기반으로 공간 이질성을 고려한 사회간접자본과 지가와의 연관성에 관한 연구

Kang Wooseok 강우석*, Kim Eunchan 김은찬**, Heo Wookjae 허욱재***

**Abstract**

An understanding of how Social Overhead Capital (SOC) is associated with the land value of the local community is important for effective urban planning. However, even within a district, there are multiple sections used for different purposes; the term for this is spatial heterogeneity. The spatial heterogeneity issue has to be considered when attempting to comprehend land prices. If there is spatial heterogeneity within a district, land prices can be managed by adopting the spatial clustering method. In this study, spatial attributes including SOC, socio-demographic features, and spatial information in a specific district are analyzed with Finite Mixture Modeling (FMM) in order to find (a) the optimal number of clusters and (b) the association among SOCs, socio-demographic features, and land prices. FMM is a tool used to find clusters and the attributes' coefficients simultaneously. Using the FMM method, the results show that four clusters exist in one district and the four clusters have different associations among SOCs, demographic features, and land prices. Policymakers and managerial administration need to look for information to make policy about land prices. The current study finds the consideration of closeness to SOC to be a significant factor on land prices and suggests the potential policy direction related to SOC.

# Ⅰ. Introduction

## 1. Social Overhead Capital and Land Price

The Social Overhead Capital (SOC) is historically categorized as an external factor of the economy (Cootner 1963;

* 국토연구원 책임연구원(제1저자) | Associate Research Fellow, Urban Research Division, Korea Research Institute for Human Settlements | Primary Author | wskang@krihs.re.kr
** 서울대학교 지능정보융합학과 박사과정수료 | Ph.D. Candidate, Department of Intelligence and Information, Seoul National University | euchan@snu.ac.kr
*** 미국 퍼듀대학교 소비자학과 조교수(교신저자) | Assistant Professor, School of Hospitality & Tourism Management, Division of Consumer Science, Purdue University | Corresponding Author | heo28@purdue.edu

Kurihara 1970). The categorization implies three points of SOC. First, that SOC is an external factor of the economy signifies that an external entity (i.e., government) owns or manages the SOCs, which are often associated with the local economy. Specifically, SOCs are social capital easily accessible to anyone in a certain area, promoting the local economy (Oxford Reference, n.d.). In other words, being owned and/or managed by a local government, SOC serves the public welfare and public well-being. Second, because they are managed by the local government, SOCs are delimited within the managerial entity's financial and economic statement. In short, the quantity and quality of SOCs in a certain area are controlled by the local government. Conversely, SOCs influence the economic growth of the local community. For instance, a local community's locations of SOCs (e.g., roads, public libraries, public health facilities, and public parks) are significant because each infrastructure and facility is associated with housing value and land price (Choi, Jeung and Park 2021). Therefore, understanding how SOCs impact the land value of the local community is necessary for effective urban planning.

## 2. Spatial Approach and Heterogeneity in Spaces

A spatial approach is needed to understand how SOCs impact land value because of urban management's complex and dynamic nature and relevant policies (Cottineau, Finance, Harna and Arcaure et al. 2019; Parr 2007). Numerous factors affect an urban area, including physical infrastructure, information, socio-economic features, and diverse cultures of the local community. Because of the complexity, spatial consideration was suggested to contribute to a more holistic understanding of urban management (Cottineau, Finance, Harna and Arcaure et al. 2019). It means that more than one rule explains the space. Therefore, various approaches to understanding an area are recommended.

For instance, even within an area, each small zone serves diverse functions in terms of infrastructure and socio-economic features (Fotheringham and Pitts 1995). Because of the variety of small zones' functions, spatial heterogeneity was introduced to explain that the linear distance from the urban center was inconsistent in many urban analytics (Oller, Martori and Madariaga 2016). Spatial heterogeneity indicates that an effect of a facility/infrastructure is not directly proportional to the distance due to other dynamic geographic factors. As a result, planning and managing an urban area's specific aspect (e.g., social overhead capital) requires spatial understanding, specifically with spatial heterogeneity. Even within one district, various characteristics exist. Therefore, a comparatively new approach to understanding a selected area is needed to investigate spatial heterogeneity.

## 3. Spatial Heterogeneity and SOCs

To follow the needs of a new approach to understanding the spatial heterogeneity, the topical issue (i.e., SOCs) has to be considered. Combining the spatial heterogeneity and the SOCs requires comprehensive considerations such as contextual attention including political, social, economic, and spatial components (Naughton 2013; Sheppard 2011). Granovetter (1985) explained social capital from a classical economic perspective, arguing that the economic conclusion may embed social relations in the shade of simplification by not providing a multi-contextual explanation. Simplifying the economic conclusion of an urban area dismisses the spatial features, such as the diversities in multiple zones, within the targeted urban area. Consequently, a categorical social capital approach was introduced and utilized (Brown and Livermore 2019; Jiao and Liu 2012; Sampson and Graif 2009). Brown and Livermore (2019) categorized the vulnerable people in Baton Rouge, LA, into four categories. Jiao and Liu (2012) found that there were hierarchical levels of clusters related to administrative districts in a Chinese province. Sampton and Graif (2009) categorized an urban area (i.e., Chicago) into four categories based on the resident-based social capital. These empirical analytic trials implied that spatial categorization may be able to serve as a solution for capturing the heterogeneity within an urban area.

## 4. Heterogeneity of Spatial Information

In terms of the important indexes of the land (i.e., price of land), heterogeneity was also a substantial issue. For instance, heterogeneity was discussed in terms of economic geography (e.g., Nocco 2009; Ottaviano 2011; Tabuchi and Thisse 2002). Tabuchi and Thisse (2002) showed that the heterogeneity of the labor force was associated with the spatial distribution of the industry. Nocco (2009) also found that the heterogeneity of the labor force (e.g., skilled and unskilled workers) was associated with the local firms' productivity. In addition, Ottaviano (2011) argued that the micro-level of heterogeneity needs to be investigated, while previous economic geography utilized macro-level heterogeneity in its survey. Even though Ottaviano's argument was intended to utilize the micro-level heterogeneity of firms and agglomeration in economics, the argument brought an idea that the micro-level heterogeneity could be an issue in land utilization and price. As a result, the issue of heterogeneity has to be considered when attempting to comprehend the land price.

Then, the following question was about how to capture heterogeneity in terms of investigating urban spaces. The approach to comprehending heterogeneity in this study was to utilize spatial clustering because the spaces can be categorized into a certain number of groups called clusters (e.g., Brown and Livermore 2019; Jiao and Liu 2012; Sampson and Graif 2009). Among multiple clusters found in a district, zones in one specific cluster were assumed to have similar attributes selected for analysis (Liu, Deng, Shi and Wang 2012). Therefore, heterogeneity might be resolved by adopting the spatial clustering method and increasing the similarities among zones in a cluster.

The research gap can be found at this point. In many of previous research in urban management, an area was investigated as a whole (e.g., Kim, Kim and Suh 2016; Suh 2005). However, as discussed above, the heterogeneity should be considered to see the complexity within a certain area. Therefore, in this study, the heterogeneity was checked so that selected factors influential effects were interpreted by having subgroups of heterogeneity. Therefore, the current study has a significant contribution to suggest a newly introduced methodology in urban management.

5. Approach with Hedonic Models

In addition to the heterogeneity issue, there is another consideration about what factors would be considered within an empirical model. The hedonic model is a well-known model to estimate a price for an environmental/ecological service directly associated with the market price (Rosen 1974). Specifically, the model is based on the assumption that the value of a product (or a service) is associated with various characteristics within the product (or the service). Therefore, the hedonic model specializes in finding an area's influential internal factors. Because of the utility of the hedonic model, it is utilized for various topics about land price, such as constructing the land price list, finding major characteristics of a city's amenities or public services, market investigation for a residential land price, and the residential locations within a selected area (Maddison 2000).

For instance, Kim, Kim and Suh (2016) found that land categories (i.e., dry paddy, building site, and community land) were associated with land price using the linear hedonic model. Specifically, the distance from a hospital was a significant factor in estimating the land price. Another study showed that population density was the major factor in land price (Suh 2005). Specifically, the number of children is strongly associated with the land price in the residential area. As such, the hedonic model helps to estimate the association between land price and the area's characteristics. This research used the characteristics of the distance to SOCs, population, female rate, and land utilities. Therefore, by using the concept of the hedonic model, land utilities and population were added.

6. Contributions

As discussed, it is important to understand the association between SOCs and land prices. However, the association is not homogeneously simple within an area. Therefore, a heterogeneous analysis needs to be performed to gain a complex understanding of an area. In addition, a hedonic model will help to find the characteristics utilized within the selected area. Therefore, this study significantly contributes to finding the association between SOCs and land price, specifically employing heterogeneous understanding. Within one district, there are multiple types of groups, and those groups might show different aspects of association among hedonic characteristics, SOCs, and land prices.

## Ⅱ. Objectives

### 1. Methodology to Solve Heterogeneity in Spatial Analysis

Even if clustering analysis is proper for finding spatial patterns, clustering analysis is weak at capturing the association among the targeted factors (e.g., the relationships among demographic features, SOCs allocations, and the land price) within each cluster. Clustering methods (e.g., k-means clustering, hierarchical clustering, density-based clustering, etc.) mostly focus on categorizing the units of analysis into a certain number of clusters (i.e., pattern finding) instead of finding the association among attributes inside the multiple clusters (Hair, Sarstedt, Ringle and Gudergan 2018; Lanza and Rhoades 2013). To estimate the association between different attributes, another step, such as linear estimation, needs to be employed after clustering analysis.

However, by employing two separate methodologies (i.e., clustering and linear estimation) as a two-step process, the measurement errors from the first analysis (clustering) are dismissed in the second analysis (linear estimation; Leisch 2004). Therefore, a set of combined methodologies between clustering and linear estimation was introduced to reduce the measurement error; these were methodologies to utilize latent variables such as latent class analysis (LCA), latent profile analysis (LPA), and finite mixture modeling (FMM). This method performed the clustering analysis with other analyses such as analysis of variance (ANOVA), ordinary least square (OLS) estimation, and logistic estimation, simultaneously (Peel and McLachlan 2000; Vermunt and Magidson 2002; Wedel and Kamakura 2000). In this study, FMM, one of the simultaneous methodologies of clustering and linear estimation (OLS), was used. In addition, the usefulness of FMM in the spatial analysis was confirmed by literature (e.g., Buddhavarapu, Scott and Prozzi 2016; Faghih-Imani and Eluru 2017). Buddhavarapu, Scott and Prozzi (2016) confirmed how the FMM captured heterogeneity appropriately by segmenting roads in a certain number of groups from the Houston area. Faghih-Imani and Eluru (2017) analyzed the city information of New York and confirmed that the FMM captured heterogeneity well in the process of spatial analysis. As such, employing FMM in spatial analysis was a feasible solution to understanding social capital's contextual consideration by handling the spatial analysis's heterogeneity issue.

### 2. Research Questions

In this study, attributes (i.e., SOCs, socio-demographic features, and spatial information) in a specific district were analyzed with FMM in order to find (a) the optimal number of clusters and (b) the association among SOCs, socio-demographic features, and the land price. The resulting research questions of the study were: RQ1 - Is there homogeneity among cells within a district or heterogeneity? RQ2 - How many categories of cells exist if there is heterogeneity? RQ3 - Do categories of the SOCs show different land values? RQ4 - Are attributes rather than SOCs (i.e., socio-economic features and spatial information) associated with land value?

To collect the data, the Architecture and Urban Research Institute's (AURI) classification by using closest facility analysis was utilized (Seong, Lim and Lim 2013). It enables the collection of spatial data for the study. The method of collecting spatial data will be introduced below. To answer the research questions above, FMM was utilized, as explained above.

## Ⅲ. Methods

### 1. Methodology

In this study, analyzed geographical data was delimited into one of the 25 districts in the metropolitan area within Gwangjin-gu, Seoul. Gwangjin-gu has a balanced combination of commercial areas and residential areas. Furthermore, as Gwangjin-gu is located on the eastern side of Seoul, adjacent to Han-River, the district has a combination of dynamics, such as an amusement park, natural park, subways, extensive transportation system,
hospitals, manufacturing areas, residential areas, etc. (https://www.gwangjin.go.kr/eng/main/main.do). Specifically, the selected district reported the highest rate of low-level buildings and the oldest buildings among 25 districts in Seoul. It implies that the selected district is the area to be investigated for better living quality by having additional SOCs.

Importing the spatial data of Gwangjin-gu according to the following methodology, this study analyzes the spatial heterogeneity of the specific area. According to the spatial data from Gwangjin-gu, 904 cells were divided by using a 100 × 100 grid polygon. <Figure 1> shows the 904 cells by the type of SOCs inside Gwangjin-gu.

**Figure 1 _ Map of SOCs Accessbility in Gwangjin-gu**

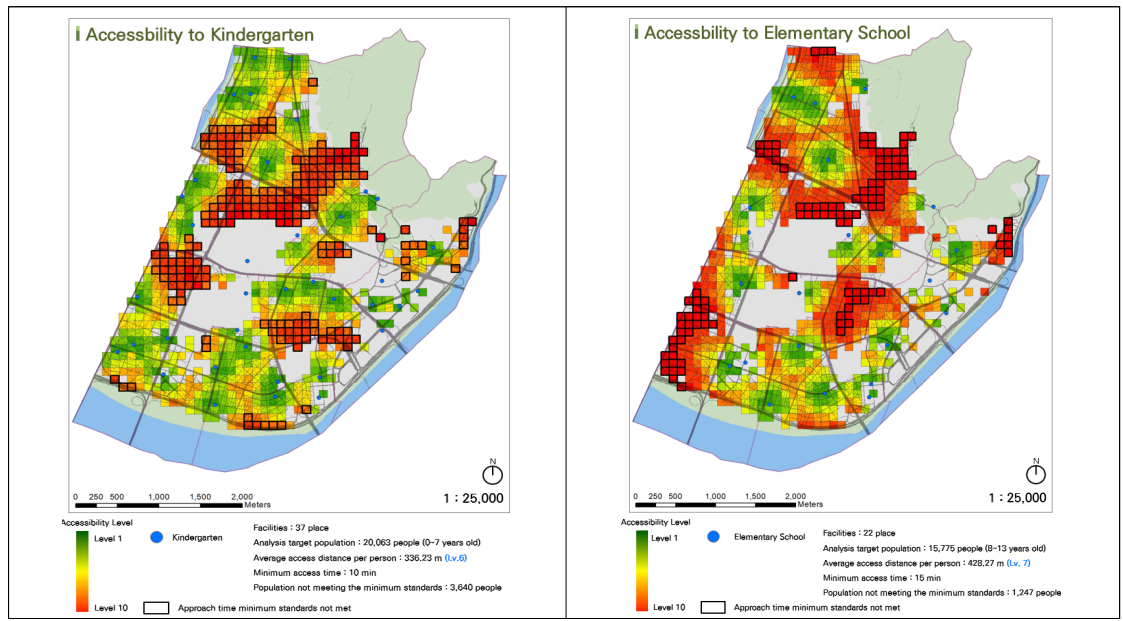

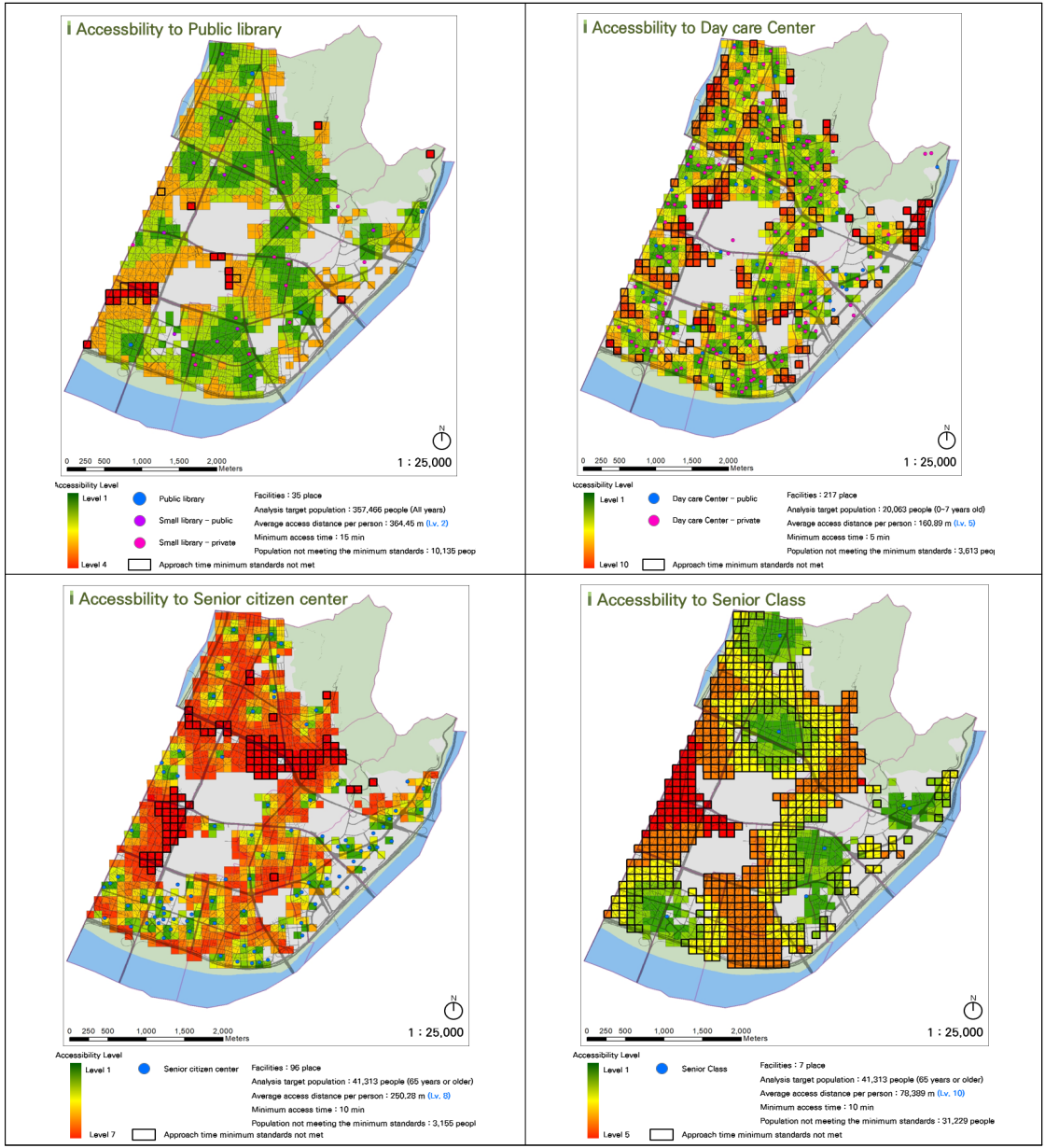

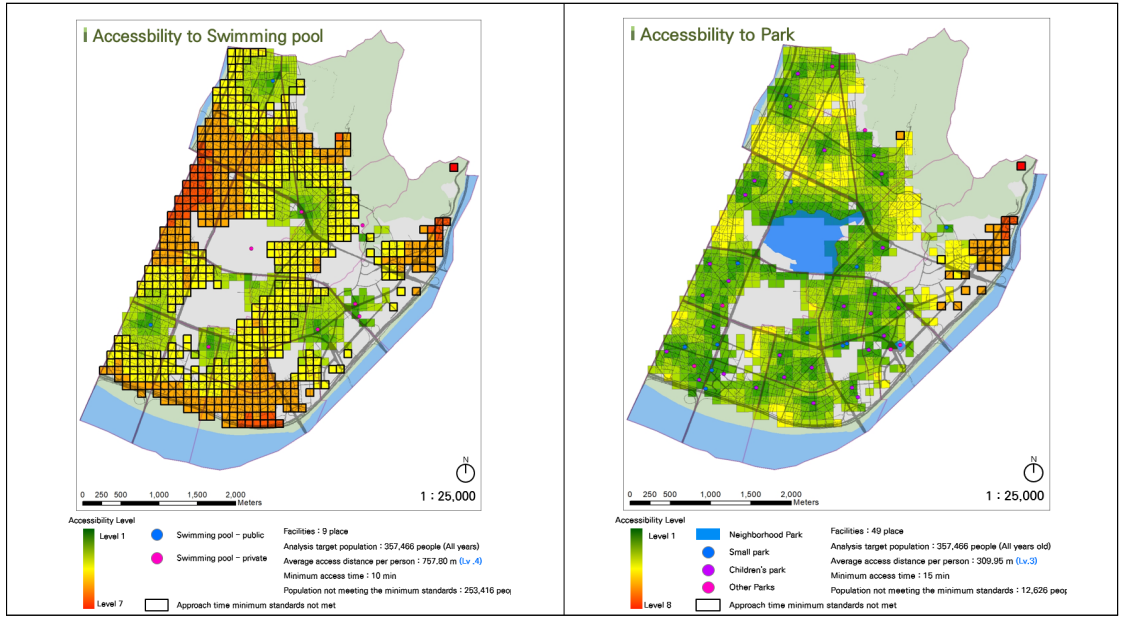

## 2. Processing of Spatial Data: National Standards for Community Infrastructure Plan

Spatial data were transformed to ranks of facilities and infrastructures by using a user-centered method. Facilities and infrastructures did not exist in all 904 cells of the target district (see <Figure 1>). This means that people in each cell have different accessibility in using the facilities and infrastructures. Therefore, the spatial data of facilities or infrastructure has to be assessed by a user-centered approach. To assess the rank of accessibility, this study utilized a two-step method.

First, the closest facility analysis (2013) was utilized, as shown in <Figure 2>. <Figure 2> shows the process of the National Standards for Community Infrastructure Plan (NSCIP) to find the closest facility. Each location of facility or infrastructure (i.e., upper line in <Figure 2>, [A]) was compared to the information of accessibility (i.e., lower line in <Figure 2>, [B]). By comparing with the lower line to check the residents' accessibility ([B] in <Figure 2>), a facility or infrastructure was selected as the closest accessible facility or infrastructure. After finding the closest facility by using [A] and [B] in the process, the walking distance ([C] in <Figure 2>) was used to calculate the distance. Therefore, even the same facility or infrastructure may have different distances in different cells in Gwangjin-gu.

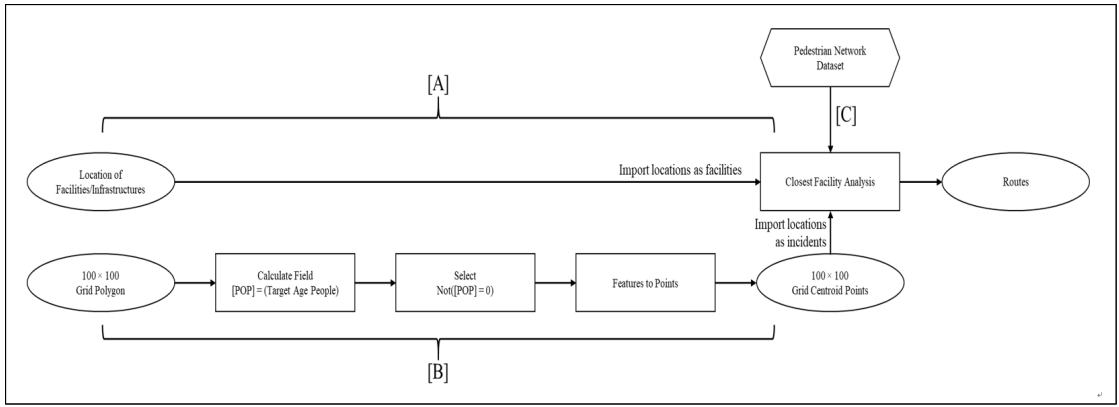

**Figure 2 _ Closest facility analysis**

Second, based on the above closest facility analysis, each facility or infrastructure was ranked by using the following threshold in <Table 1>. Based on the guideline for assessing the basic infrastructure for a living (Ministry of Land, Infrastructure, and Transport [MOLIT] 2019), each facility was assessed by using the time to reach the facility on foot. For instance, for the first grade, the minimum time to reach a kindergarten was suggested as much as 5 to 10 minutes on foot; elementary school was suggested as much as 10 to 15 minutes; and health facilities were suggested as much as 10 minutes. Based on the criteria for assessing each facility's accessibility by MOLIT (2019), the average distance of each grade was shown in <Table 1>. For instance, if the average distance to a kindergarten from a specific cell (i.e., the closest cell from the facility) was assessed as 250 meters, then the grade of the kindergarten was assessed as 4th grade for people living in the specific cell. In summary, the first step (<Figure 2>) finds the closest facility to each cell in Gwangjin-gu and the second step (<Table 1>) decides the grade based on the length of each cell.

**Table 1 _ Grades of Facilities by Distances**

(unit: meter)

| Grade | Kind. | Ele. Sch. | Lib. | Day C. | Sn. Comm. | Sn. Ed. | Hth. Faci. | Parks |
|---|---|---|---|---|---|---|---|---|
| 1 | 128 | 154 | 253 | 71 | 58 | 378 | 150 | 156 |
| 2 | 184 | 205 | 448 | 96 | 75 | 628 | 280 | 265 |
| 3 | 233 | 253 | 758 | 121 | 92 | 953 | 518 | 438 |
| 4 | 283 | 302 | 1,275 | 148 | 112 | 1,449 | 932 | 761 |
| 5 | 335 | 351 | 1,909 | 178 | 137 | 2,217 | 1,481 | 1,266 |
| 6 | 395 | 405 | 2,637 | 213 | 169 | 3,369 | 2,163 | 1,914 |
| 7 | 468 | 471 | 3,494 | 257 | 215 | 5,352 | 3,006 | 2,734 |
| 8 | 571 | 561 | 4,625 | 312 | 289 | 8,465 | 4,146 | 3,845 |
| 9 | 771 | 731 | 6,522 | 404 | 49 | 13,386 | 6,169 | 5,656 |
| 10 | 17,116 | 5,897 | 27,753 | 19,632 | 9,486 | 87,815 | 28,088 | 20,627 |

Note: The numbers indicate the maximum distance (i.e., threshold) for the grade. Kind. = kindergarten; Ele. Sch. = elementary schools; Lib. = library; Day C. = day care; Sn. Comm. = senior community; Sn. Ed. = senior educations; Hth. Faci. = health facilities; Parks = knighthood park and public park

## 3. Empirical Model

In the empirical model, the land value and population were measured with a logarithms format because each cell has a different density in population. In other words, the land price was weighted by population in all areas (i.e., 904 cells within the same district). Because the study is concerned about the distance between people and the social facilities within each cell, population density needs to be considered to investigate the land price in urban areas (Glumac, Herrera-Gomez and Licheron 2019; Dong, Li, Zhang and Di 2016). Specifically, Dong, Li, Zhang and Di (2016) suggested the population-weighted efficiency for managing the selected area. Their suggestion was to consider the population as a denominator of the distance across selected local points. Specifically, the land price is associated with the demand for purchasing land. The demand for land is also associated with the population living in the selected area. Therefore, to consider the population density of a cell, the ratio between land price and the population was initially considered by using the logarithms format.

To adopt the population in the empirical modeling, the logarithm of land price was utilized as the dependent variable of the empirical model and the logarithm format of the population is utilized as control variable in the empirical model. By using the logarithms format, the association between land price and the population was consistent between the population- weighted efficiency (i.e., <Equation 1>) and the empirical modeling (i.e., <Equation 2>). As result, by considering that the dependent variable of the study—the land value—the basic format of the association between land value and the population was <Equation 1>.

$$\ln\left(\frac{LandPrice}{Population}\right) = \ln(LandPrice) - \ln(Population) \ldots \text{Equation 1.}$$

This study assumed that there is heterogeneity in explaining land value among 904 cells within the same district in Seoul. In other words, 904 cells are assumed to be categorized as subgroups when considering the association between SOC facilities and land value. To check whether SOCs and basic information about the land were associated with land value, the following formula (<Equation 2>) was utilized as the basic model.

$$\ln(LandPrice) = a + \beta_i Grade_i + \beta_p \ln(Population)_p + \beta_j Control_j + e$$

…Equation 2.

where, a is the constant; $\beta$ i denotes the coefficients of SOCs (i.e., kindergarten, daycare, elementary school, public library, senior community, senior education, health facilities, neighborhood park, and public park); $\beta$ p signifies the coefficient for the logarithm format of population; $\beta$ j is the coefficient for control variables (i.e., female rate in a cell, public land rate in a cell, commercial land rate in a cell, and green land rate in a cell). The empirical model is used as the basic format to assess how SOCs affect the land value.

## 4. Spatial Heterogeneity: Finite Mixture Model

As explained above, the main concern of the study was to assess heterogeneity (RQ1 and RQ2) and to see how SOCs were associated with land value in each sub-group (RQ3 and RQ4). To answer the research questions as mentioned above, FMM estimates the sub-groups by using <Equation 3> below (StataCorp 2021):

$$p(\ln(LandPrice)|\Theta) = \sum_{k=1}^{K} a_k p_k(\ln(LandPrice)|X_i = k, \beta_k) \quad \ldots \text{Equation 3.}$$

Here, $p_k$ (ln (LandPrice) | $X_i$=k,$\theta_k$) denotes the kth component density with coefficients $\beta_k$ (i.e., $\beta_i, \beta_p$ and $\beta_j$); p(ln (LandPrice) | $\Theta$) denotes the discrete dependent variable (i.e., ln(LandPrice)) with its probability distribution; $X_i$ would be the k-dimensional factors (i.e., variables in <Equation 2>); $a_k$ denotes the marginal distribution of the kth component. As shown in <Equation 3>, the basic model from <Equation 2> was assumed to be categorized into sub-groups (i.e., k).

## 5. Empirical Data

The following spatial data was extracted from each administrative division within Gwangjin-gu by using a geocoding program: (a) division of construction; (b) division of real estate; (c) division of urban planning; and (d) division of finance. Specifically, local addresses were recoded to coordinates on the map (see <Figure 1>) by using ArcMap 10.3 and GeocoderXR. The foundational coordinate was Korea 2000 Unified Coordinate System with 100 square meters grid. In Gwanjin-gu, Seoul, there were 904 cells by using 100 square meters. Aggregated information of residences of each cell was utilized as the unit of analysis in this study. The descriptive information of 904 cells (e.g., population, income, female rate, land price, etc.) are presented in the next result.

## 6. Measurements

The dependent variable is the land value for the logarithm format for land price. From the spatial data collected for this study, the land price of a cell was determined by the average price of all land within the cell. The measurement unit was denoted in Korean currency (Won). Originally, the ratio between price and population was the targeted measurement in this study.
As shown on <Equation 1> and <Equation 2>, the ratio between price and population can be converted to a logarithm format with a linear estimation. As a result, the dependent variable is the logarithm format of price, which is assumed to be associated with the logarithm format of population.

The major independent variables for the estimation were population and various SOCs, including kindergartens,

elementary schools, public libraries, daycare centers, community centers for senior citizens, continuing education centers for senior citizens, health-related facilities (i.e., swimming pool, physical education center, fitness center), neighborhood parks, and public parks. The population of a cell was recorded as the total number of residents in each cell, and the population was transformed to a logarithm format (see <Equation 2>). The SOCs were coded with the NSCIP as explained above. Each SOC was expressed on a scale from 1 to 11 based on the NSCIP (see <Table 1>). The smallest number (= 1) denotes that a SOC facility is located at a place where people in the cell are able to access the facility in the shortest distance; vice-versa, the largest number (= 11) means the longest distance. All the SOCs were transformed to standardized scores to unify the scales.

As the control variables, features of cells were included in the estimating model: the proportion of females in a cell, the proportion of public land in a cell, and the proportion of green land in a cell. The proportion of females in a cell was calculated by the ratio between female numbers and the total population in the cell. The proportion of public land in a cell was measured by the ratio between public land and the total land size in the cell. The proportion of green land in a cell was assessed by the ratio between green land and the total land size in the cell.

## Ⅳ. Result

### 1. Descriptive Information of Samples

As shown in <Table 2>, the average land price of the 904 cells was 110,000,000 won, with a standard deviation of 67,500,000 won (per cell). The most expensive land price was approximately 478,000,000 won. The NSCIP of SOCs showed different averages and standard deviations. The average population across 904 cells was 395.43 people with the standard deviation of 504.76. The average female rate among 904 cells was 50% with standard deviation of 0.08. Across 904 cells, the proportion of public land was 29% compared to private land. Compared to residential areas, green land had the proportion of 2%, and commercial areas had the proportion of 4%. The detail of the descriptive information was shown in <Table 2>.

Table 2 _ Descriptive Information of 904 Cells

|  | Mean | S.D. | Min. | Max. |
|---|---|---|---|---|
| Land price (1 million won) | 110 | 6.75 | 0 | 47.8 |
| SOCs |  |  |  |  |
| Kindergarten | 6.06 | 2.91 | 1.00 | 11.00 |
| Elementary school | 7.03 | 2.89 | 1.00 | 11.00 |
| Public library | 2.02 | 0.83 | 1.00 | 11.00 |
| Daycare | 5.42 | 2.98 | 1.00 | 11.00 |
| Senior community | 7.39 | 2.52 | 1.00 | 11.00 |
| Senior education | 3.19 | 1.86 | 1.00 | 11.00 |
| Health facilities | 8.26 | 1.89 | 3.00 | 11.00 |

|  | Mean | S.D. | Min. | Max. |
|---|---|---|---|---|
| Neighborhood park | 2.54 | 1.11 | 1.00 | 11.00 |
| Public park | 4.51 | 1.54 | 1.00 | 11.00 |
| Demographic Factors |  |  |  |  |
| Population | 395.43 | 504.76 | 1 | 10161 |
| Female rate | 0.50 | 0.08 | 0 | 1.00 |
| Usage of Land |  |  |  |  |
| Rate of public land | 0.29 | 0.23 | 0 | 1.00 |
| Green land | 0.02 | 0.08 | 0 | 0.61 |
| Commercials | 0.04 | 0.16 | 0 | 1.00 |

## 2. Model Selection: Existence of Heterogeneity of the Analyzed Area

Based on the assumption that 904 cells are not homogeneous, but heterogeneous, because of the geographical diversity in the district, multiple sub-groups were expected to exist. <Table 3> shows the statistically optimal numbers of sub-groups found by FMM. Generally, Akaike Information Criteria (AIC) and the Bayesian Information Criteria (BIC) were expected to have a low optimal number of sub-groups (Hair, Sarstedt, Ringle and Gudergan 2018; Lanza and Rhoades 2013; Nest, Passos, Candel and Breukelen 2020; Sarstedt 2008). In terms of Normalized Entropy Criterion (NEC), the smallest numbers were recommended to select the optimal number of sub-groups (Biernacki, Celeux and Govaert 1999; Nest, Passos, Candel and Breukelen 2020). As a result, the optimal number of sub-groups was 4.

Table 3 _ Model Selection

|  | Log likelihood | df | AIC | BIC | NEC |
|---|---|---|---|---|---|
| 2 Groups | -823.82 | 33 | 1713.65 | 1872.24 | .6659 |
| 3 Groups | -790.27 | 50 | 1680.54 | 1920.82 | .5900 |
| 4 Groups | -738.94 | 67 | 1611.88 | 1933.86 | .3822 |
| 5 Groups | -668.35 | 84 | 1504.69 | 1908.37 | .6845 |
| 6 Groups | -578.47 | 101 | 1358.93 | 1844.31 | .7547 |

Examination of the model selection answered RQ1 and RQ2. The optimal number of sub-groups was four (RQ2). This implies that four groups of cells in the district are the best categorization rather than the other number of groups., which was possibly 3. However, NEC of 3 groups is comparatively higher than 4 groups, which meant that the usage of 3 groups might bring the measurement error higher than 4 groups. In terms of 2 groups and 5 groups, their NECs are much higher than 4 groups. Lastly, 6 groups showed the least acceptable range in terms of NEC and AIC. As a result, the cells from different types showed different features, which means heterogeneity (RQ1).

### 3. Descriptive Information of Sub-groups

<Table 4> shows the descriptive information of four groups among the 904 cells, and <Figure 3> shows the geographical location of four groups. The first group consists of 389 (40.03%) cells; the second group consists of 265 (29.31%) cells; the third group consists of 144 (15.93%) cells; and, finally, the fourth group consists of 106 (11.73%) cells.

Geographically, by examining <Table 4> and <Figure 3>, Group 1 was more likely to be a residential areas, but Groups 3 and 4 were more likely to be a complex area. Group 2 was interpreted to be an intermediate area between residential and complex areas. This was confirmed by the descriptive explanation of <Table 4>. Group 1 had longer distances to elementary schools and senior community centers but shorter distances from other SOCs; Groups 3 and 4 showed a shorter distance from elementary schools and senior community centers but longer distances to most other SOCs; and Group 2 showed the mixed features of Groups 1, 3 and 4.

Second, in the case of land usage, Groups 1 and 2 were more likely to be residential areas, but Groups 3 and 4 were more likely to be complex areas. For instance, Groups 1 and 2 showed the rate of public land as much as 24% (SD = 16%) and 29% (SD = 19%), respectively; but, Groups 3 and 4 showed the rate of public land as high as 38% (SD = 30%) and 35% (SD = 34%), respectively. In addition, the rate of commercials in Groups 1 and 2 were as low as 0% (SD = 3%) and 3% (SD = 13%), respectively; but, the rate of commercials in Groups 3 and 4 were comparatively high at 10% (SD = 25%) and 8% (SD = 25%), respectively.

Table 4 _ Descriptive Information of 904 Cells by Sub-groups

|  | Group 1 (n = 389) Mean (S.D.) | Group 2 (n = 265) Mean (S.D.) | Group 3 (n = 144) Mean (S.D.) | Group 4 (n = 106) Mean (S.D.) |
|---|---|---|---|---|
| Land price (1 million won) | 13.4 (4.31) | 13 (6.32) | 5.88 (4.91) | 3.98 (8.53) |
| SOCs |  |  |  |  |
| Kindergarten | -.03 (1.00) | -.07 (.96) | .15 (.97) | .08 (1.09) |
| Daycare | -.11 (.98) | -.08 (.93) | .17 (1.07) | .40 (1.03) |
| Elementary school | .05 (.98) | -.04 (.98) | -.03 (1.02) | -.03 (1.11) |
| Public library | -.14 (.96) | -.08 (.99) | .24 (1.01) | .38 (1.01) |
| Senior community | .09 (.90) | -.04 (1.01) | -.20 (1.12) | .03 (1.11) |
| Senior education | -.05 (.97) | -.11 (.62) | .06 (1.00) | .36 (1.62) |
| Health facilities | -.05 (.91) | -.06 (1.05) | .10 (1.11) | .20 (1.01) |
| Neighborhood park | -.02 (.94) | -.14 (.98) | .29 (1.11) | .04 (1.03) |
| Public park | -.08 (.97) | .07 (.93) | .11 (1.03) | -.01 (1.22) |
| Population | 5.60 (1.09) | 5.58 (1.07) | 5.34 (1.39) | 4.69 (2.12) |
| Female rate | .50 (.07) | .51 (.05) | .51 (.07) | .50 (.15) |
| Rate of public land | .24 (.16) | .29 (.19) | .38 (.30) | .35 (.34) |
| Land Utilities |  |  |  |  |
| Green land | .01 (.07) | .02 (.10) | .03 (.10) | .01 (.04) |
| Commercials | .00 (.03) | .03 (.13) | .10 (.25) | .08 (.25) |

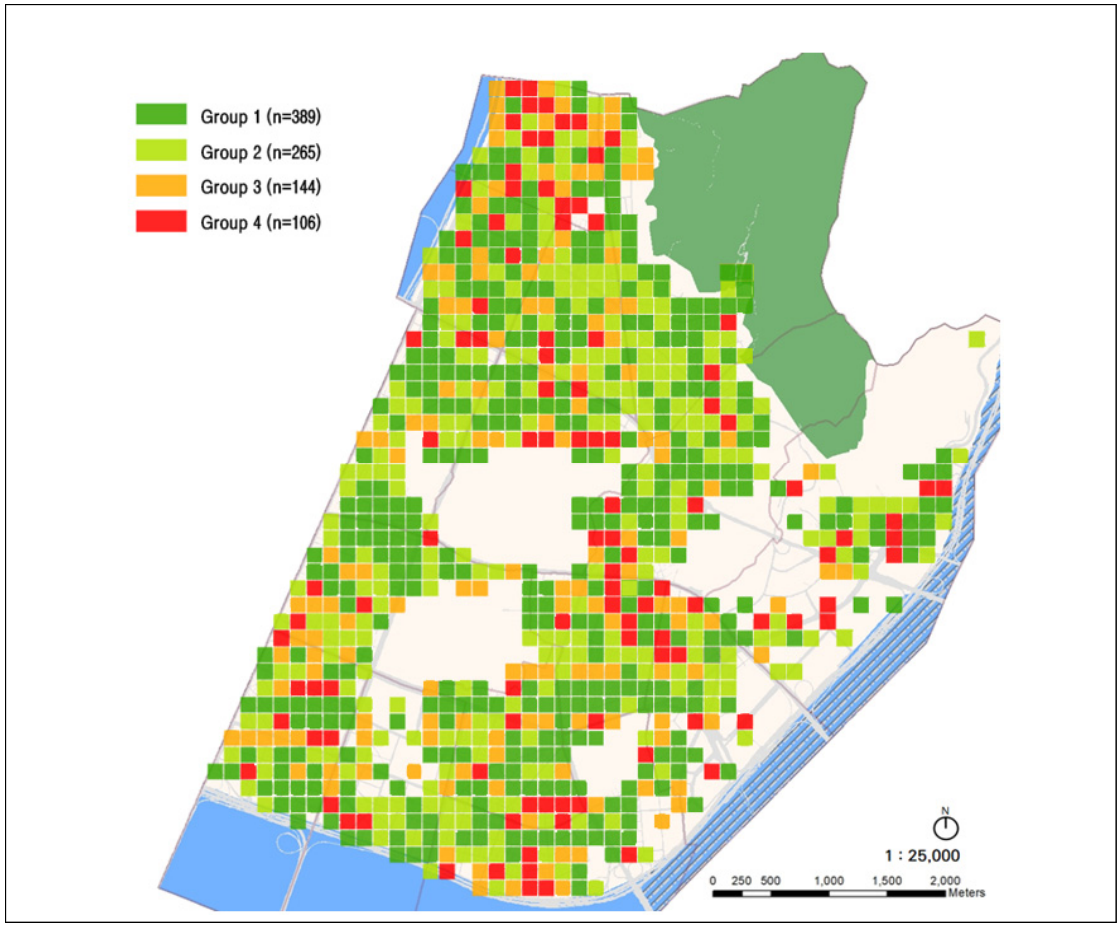

Figure 3 _ 4 Sub-groups in Gwangjin-gu

However, the green area rates in all four groups were similar, around 1% to 3%. Therefore, based on land price, population, the proportion of female citizens, public land, and land utilities, Groups 1 and 2 are considered residential areas, and Groups 3 and 4 are interpreted to be complex or commercial areas (see <Appendix>). The table in <Appendix> is the text summary of <Table 4>, which is the same information but promotes a better understanding of <Table 4>.

Third, there are different characteristics of SOCs shown by each group. Regarding facilities for younger children (i.e., kindergarten and daycare), Groups 1 and 2 showed shorter distances to SOCs than Group 3. Considering that the distance to SOCs was coded as a standardized value, the negative values of Groups 1 and 2 indicated shorter distances to SOCs. However, the positive value in Groups 3 and 4 indicated longer distances to SOCs. For instance, Group 1 showed -.03 (SD = 1.00) for kindergartens and -.11 (SD = .98) for daycare centers; Group 2 showed -.07 (SD = .96) for Kindergarten and -.08 (SD = .93) for daycare centers; Group 3 showed .15 (SD = .97) for kindergartens and .17 (SD = .98) for daycare centers; and Group 4 showed .08 (SD = 1.09) for kindergartens and .40 (SD = 1.03) for daycare centers.

Fourth, in terms of educational SOCs (i.e., elementary schools and public libraries), Group 1 and the other three groups (i.e., Groups 2, 3, and 4) showed opposite patterns. Group 1 was further away from school but closer to public libraries. Elementary schools were further away from residents (M = .05, SD = .98) in Group 1; but schools were located close to residents in Groups 2, 3, and 4 (M = -.04, SD = .98; M = -.03, SD = 1.02; M = -.03, SD = 1.11, respectively). Public libraries were closer to residents in Groups 1 and 2 (M = -.14, SD = .96; M = -.08, SD = .99, respectively); but public libraries were further away from residents in Groups 3 and 4 (M = .24, SD = 1.01; M = .38, SD = 1.01, respectively).

Fifth, regarding SOCs for senior citizens (i.e., senior community and senior education), these SOCs were similar to the case of educational facilities. Group 1 and the other three groups (i.e., Groups 2, 3, and 4) showed opposite patterns. Senior community was far from residents in Groups 1 and 4 (M = .09, SD = .90; M = .03, SD = 1.11, respectively); but communities were located close to residents in Groups 2 and 3 (M = -.04, SD = 1.01; M = -.20, SD = 1.12, respectively). Senior education facilities were close to residents in Groups 1 and 2 (M = -.05, SD = .97; M = -.11, SD = .62, respectively); but education centers were far from residents in Groups 3 and 4 (M = .06, SD = 1.00; M = .36, SD = 1.62, respectively).

Sixth, Groups 1 and 2 had short distances between residents and health facilities and parks. However, Groups 3 and 4 had long distances between residents and health facilities and parks. Group 1 showed short distances from health facilities, neighborhood parks, and public parks (M = -.05, SD = .91; M = -.05, SD = .94; M = -.08, SD = .97, respectively); Group 2 showed short distances from health facilities and neighborhood parks but long distances from public parks (M = -.06, SD = 1.05; M = -.14, SD = .98; M = .07, SD = .93, respectively); Group 3 showed long distances from health facilities, neighborhood parks, and public parks (M = .10, SD = 1.11; M = .29, SD = 1.11; M = .11, SD = 1.03, respectively); and Group 4 showed long distances from health facilities and neighborhood parks but short distance from public parks (M = .20, SD = 1.01; M = .04, SD = 1.03; M = -.01, SD = 1.22, respectively).

To sum up the descriptive information, Groups 1 and 2 are residential areas with short distances to SOCs and Groups 3 and 4 are commercial or complex areas with long distances to SOCs. Among Groups 1 and 2, Group 2 showed a shorter length to SOCs compared to Group 1. Between Groups 3 and 4, Group 4 showed longer distances to SOCs compared to Group 3.

## 4. Land Price by Sub-groups' SOCs

<Table 5> shows the marginal effects of SOCs on land price. As explained above in <Equation 1> and <Equation 3>, FMM executed clustering algorithms, implementing multiple regressions simultaneously. The dependent variable of all multiple regressions was the logarithm format of land price. As shown in <Equation 2>, the logarithms were used for controlling the effect of population density.

If all 904 cells were to be homogeneous, the first column of <Table 5> would explain the marginal effects of the distances to SOCs on the land price (i.e., model with total 904 cells). Based on an assumption of homogeneity, the distances of several SOCs were positively associated with land price, such as elementary school (b = .11, p < .01) and senior community centers (b = .19, p < .001); and, distance of a few other SOCs such as public libraries (b = -.13, p < .001) and neighborhood parks (b = -.14, p < .001) were negatively associated with land prices. This means that, if the selected area (Gwangjin-gu) is homogenous about the association between SOCs and land price, the easy accessibility to elementary school and senior community centers increased the land price but the public library and neighborhood park decreased the land price. Considering that the land price is generally associated with the residents' demand for land, the homogeneous assumption explains that Gwanjin-gu's residents prefer the educational and senior SOCs but avoid the recreational SOCs. Based on that SOCs are built and managed for the quality of life of the residents, the results from homogeneous assumption brings abnormal results. Therefore, the homogeneous assumption needs to be checked first.

On the other hand, the assumption that 904 cells are not homogeneous (i.e., heterogeneous), the association between land price and SOCs were different with the total model, such as <Table 5> and the following explanation.

Table 5 _ Finite Mixture Modeling Regression Results by Subgroups

|  | Total (n = 904) b (S.E.) | Group 1 (n = 389) b (S.E.) | Group 2 (n = 265) b (S.E.) | Group 3 (n = 144) b (S.E.) | Group 4 (n = 106) b (S.E.) |
|---|---|---|---|---|---|
| SOCs |  |  |  |  |  |
| Kindergarten | .01 (.04) | -.07** (.02) | .04 (.02) | -.02 (.04) | .06 (.21) |
| Elementary school | .11** (.04) | .03 (.02) | -.09** (.02) | .05 (.04) | .43* (.18) |
| Public library | -.13*** (.04) | .02 (.02) | .04 (.02) | -.09* (.04) | -.62*** (.18) |
| Daycare | -.06 (.04) | .02 (.02) | .02 (.03) | -.07* (.04) | .00 (.19) |
| Senior community | .19*** (.04) | -.03 (.02) | .14*** (.02) | .39*** (.04) | .37* (.18) |
| Senior education | -.08 (.04) | .02 (.02) | .22*** (.03) | .08* (.04) | -.23 (.15) |
| Health facilities | -.05 (.04) | -.07** (.02) | -.03 (.02) | .02 (.04) | -.26 (.18) |
| Neighborhood park | -.14*** (.04) | -.04 (.02) | -.07*** (.02) | -.23*** (.04) | -.13 (.19) |
| Public park | .07 (.04) | -.02 (.02) | .04 (.03) | .12** (.04) | .24 (.16) |

|  | Total (n = 904) b (S.E.) | Group 1 (n = 389) b (S.E.) | Group 2 (n = 265) b (S.E.) | Group 3 (n = 144) b (S.E.) | Group 4 (n = 106) b (S.E.) |
|---|---|---|---|---|---|
| Population | .16*** (.04) | .14*** (.03) | -.07** (.02) | -.08** (.03) | .22 (.12) |
| Female rate | -1.34** (.43) | -.46 (.26) | .33 (.37) | -.64 (.47) | -1.40 (1.10) |
| Rate of public land | -1.70*** (.17) | -1.06*** (.13) | -2.18*** (.18) | -2.24*** (.13) | -1.05* (.52) |
| Land Utilities |  |  |  |  |  |
| Commercials | 1.77*** (.42) | 1.87*** (.26) | 1.02*** (.19) | .63 (.34) | 2.62 (2.35) |
| Green land | -2.23*** (.25) | -6.58*** (.55) | -1.56*** (.16) | -2.54*** (.16) | -1.14 (.74) |
| Constant | 18.45*** (.30) | 18.33*** (.19) | 19.42*** (.21) | 19.55*** (.34) | 16.71*** (.91) |
| Margin |  | .34 | .29 | .21 | .16 |
| S.E. |  | .04 | .04 | .03 | .02 |
| F | 41.05*** |  |  |  |  |
| $R^2$ | .39 |  |  |  |  |

Nore : * p < .05; ** p < .01; *** p < .001. Constant denotes the intercept from the regression result.

In Group 1, the distance of two SOCs were found to be negatively associated with land price, kindergarten (b = -.07, p < .01), and health facilities (b = -.07, p < .01). The shorter distances to kindergartens and health facilities cause the higher price of land in the cells. This implies that cells in Group 1 did not show any association between land price and distance to other SOCs (i.e., elementary schools, public libraries, daycare centers, senior community centers, senior education centers, neighborhood parks, and public parks). Considering that the land price is generally associated with the demand for land, Group 1 residents do not care about the accessibility to SOCs. Even easy accessibility to SOCs did not increase nor decrease the land price.

In Group 2, the distances to four SOCs were found to be significantly associated with land price: the distances to elementary schools and neighborhood parks were negatively associated with land price (b = -.09, p < .01; b = -.07, p < .001, respectively); but, the distances to senior community centers and senior education centers were positively associated with land price (b = .14, p < .001; b = .22, p < .001, respectively). The shorter distance to elementary school and neighborhood park caused the higher price of land in the cells. However, the longer distance to senior SOCs led to a higher price of land in the cells. Considering that the land price is generally associated with the demand for land, in Group 2, residents prefer the educational SOC and recreational SOC but avoid the senior facilities.

In Group 3, the distances to six SOCs were found to be significantly associated with land price: public libraries, daycare centers, and neighborhood parks were negatively associated with land price (b = -.09, p < .05; b = -.07, p < .05; b = -.23, p < .001, respectively); but, senior community centers, senior education centers, and public park were positively associated with land price (b = .39, p < .001; b = .08, p < .05; b = .12, p < .01, respectively). https://www.moomooz.co.kr/goods/goods_list.php?page=2&brandCd=B00000IZThe shorter distances to public libraries, daycare centers, and neighborhood parks led to a higher price of land in the cells. However, the longer distances to senior SOCs and public parks led to a higher price of land in the cells. In Group 3, residents showed a similar tendency of showing preference to educational and recreational SOCs like Group 2. However, the residents in Group 3 showed more preference on educational SOCs.

In Group 4, the distances of three SOCs were found to be significantly associated with land price: public library was negatively associated with land price (b = -.62, p < .001); but elementary schools and senior community centers were positively associated with land price (b = .43, p < .05; b = .37, p < .05, respectively). The shorter distance to public library led to a higher price of land in cells.

However, the longer distances to elementary schools and senior community centers meant a higher price of land in the cells. Considering that the land price is generally associated with the demand for land, residents in Group 4 showed totally opposite tendency compared to Groups 2 and 3. Residents in Groups 2 and 3 showed a preference for educational and recreational SOCs, but residents in Group 4 showed avoidance of educational SOCs. This implies that residents in Group 4 were not associated with young children in their lifestyle. Specifically, considering that the cells in Group 4 have a higher rate of commercial areas (25%), the educational needs can be lowered in Group 4 compared to other groups.

Finally, the socio-demographic factors also showed different associations with land price across four groups. Population was positively associated with land price in Group 1 (b = .14; p < .001) but negatively associated with land price in Groups 2 and 3 (b = -.07, p < .01; b = -.08, p < .01, respectively). The proportion of female citizens was significant in the total model (b = -1.34, p < .01) but not significant in each of the four groups.

In all groups, the proportion of public land caused the reduction of land price (b = -1.06, p < .001; b = -2.18, p < .001; b = -2.24, p < .001; b = -1.05, p < .05, respectively). In Groups 1 and 2, the rate of commercials increased the land price (b = 1.87, p < .001; b = 1.02, p < .001, respectively).

However, in Groups 1, 2, and 3, the green land caused a reduction of land price (b = -6.58, p < .001; b = -1.56, p < .001; b = -2.54, p < .001, respectively). As a result, based on the heterogeneity assumption of 904 cells, four groups showed different associations between SOCs, socio-demographic factors, and land price.

All results above were not problematic with multicollinearity, as shown in <Table 6>. As shown in <Table 6>, the variance inflation factor (vif) of the total model was 1.06, which was lower than the suggested threshold of 5.0 (Simon 2009). In addition, all vif's of all independent variables were lower than 5.0. Therefore, the results from FMM were considered results without the issue of multicollinearity.

Table 6 _ Multicollinearity Test

| | vif | 1/vif |
|---|---|---|
| SOCs | | |
| Kindergarten | 1.47 | .68 |
| Elementary school | 1.26 | .79 |
| Public library | 1.15 | .87 |
| Daycare | 1.54 | .65 |
| Senior community | 1.19 | .84 |
| Senior education | 1.46 | .69 |
| Health facilities | 1.19 | .84 |
| Neighborhood park | 1.26 | .79 |
| Public park | 1.19 | .84 |
| Population | 1.90 | .53 |
| Female rate | 1.06 | .94 |
| Rate of public land | 1.30 | .77 |
| Land Utilities | | |
| Commercials | 1.09 | .92 |
| Green land | 1.36 | .74 |
| Total | 1.06 | .94 |

## V. Discussion and Conclusion

To consider the feature of heterogeneity in urban management, this study utilized a method of FMM. The strength of the FMM was coming from its mathematical soundness, which produced the interpretability (McLachlan 2009). In addition, the mathematical estimation of clustering and consequent regression occurred simultaneously so that the estimation procedure minimized the measurement error (Peel and McLachlan 2000; Vermunt and Magidson 2002; Wedel and Kamakura 2000). Finally, the strength of the FMM is to produce interpretable coefficients of the selected factors with having subgroups. Furthermore, by accepting the theoretical concept of hedonic model inside of FMM, the interpretable coefficients of the selected factors were more closed to theoretical explanation. Therefore, the feature of heterogeneity within a certain area can be more captured by using FMM like below discussions about subgroups.

In terms of Group 1, better accessibility to kindergartens and health facilities was associated with higher land prices. However, considering that other SOCs were not significantly associated with land price, people residing in Group 1 areas did not impact SOCs greatly. Instead, it was necessary to check two kinds of results in Group 1: (a) the positively significant association among the population, commercial areas, and land price; and (b) the negatively significant association among green land, public land, and land price. Based on the two sets of results, people residing in the Group 1 area were more likely to have been concerned about the commercialized infrastructure or commercialized utilization. Considering that cells in Group 1 were residential areas, this implies that people in Group 1 were possibly concerned with land value instead of living quality. Based on the findings, the demand for the land or needs for the land is not concerned about SOCs in Group 1. It implies that the government does not need to spend significant resources on the area. However, further investigation such as behavioral response to the land usage among the residents within the cells of Group 1 may be needed, which may help policymakers promote, suspend, and control the price changes of the land within the cells.

Group 2 showed different results than Group 1, even though both groups are residential areas. In Group 2, better accessibility to elementary schools and neighborhood parks was associated with higher land prices. In addition, lower population and public usage were associated with higher land prices. It implies that the residents living in cells of Group 2 are concerned about the quality of life. Interestingly, the inferior accessibility to senior community and education centers was associated with higher land prices. It might indicate that the people in Group 2 were more likely to be separated from older generations. However, commercial utilities still increased the land price, as observed in the case of Group 1. It implies that the essential commercial areas increased the land price regardless of the residents. Governing entities may have an issue with the Group 2 cells because the demand for land is more complex (i.e., a combination of living and commercial). However, based on the findings that "young" and "commercial" lead the land price, governing entities may choose the better SOCs to improve the living quality in the cells of Group 2.

Groups 3 and 4 are not residential areas but complex areas, as noted in the above results section. Therefore, cells in Groups 3 and 4 showed a different association between SOCs and land price. In terms of Group 3, better accessibility to public libraries, daycare centers, and neighborhood parks was associated with higher land prices, but inferior accessibility to senior communities, senior education centers, and public parks was associated with higher land prices. The features found in Group 3 were somewhat similar to the features of Group 2. However, as noted above, Group 3 is more likely to be a complex area including commercial utilities different from the features of residential areas of Group 2. This implies that people living in the cells of Group 3 are more likely to live near their workplace. People in cells of Group 3 cared about the SOCs as well as the land usage. Based on the findings about Group 3, governing entities can understand the demand for land in the cells. For instance, more educational SOCs than Group 2 are needed within Group 3 cells. However, similar range of recreational SOCs compared to Group 2 are recommended to the cells in Group 3.

Compared to the other three groups, cells in Group 4 showed that this area was less associated with quality of life. For instance, the better accessibility of a public library was the only one that was positively associated with higher land prices, but the inferior accessibility to senior community centers and elementary schools was associated with higher land prices. As for complex areas, Group 4 showed that people living in these cells were less concerned with the overall distances to other SOCs. Seeing the low demographic density of the cells in Group 4 and the higher rate of commercial area, cells in Group 4 were more likely to require recreational SOCs instead of other SOCs.

This study investigated the overall possibility of spatial heterogeneity within the same district. Because the study reveals the potential existence of spatial heterogeneity in one case (i.e., Gwangjin-gu), deeper investigations including behavioral factors and governmental issues are recommended for further study. The results and discussions bring a serious consideration of SOCs in managerial techniques of land price in South Korea. Because the land price including housing value continue to plague South Korea (Kim 2021), policymakers and managerial administration needs to look for appropriate information to make policies about the land prices. Based on the current study, the consideration of SOCs is one of the significant factors on the land prices.

However, SOCs are not always related to land price, as shown in <Table 5>. Some cells, such as the 144 cells in Group 3, showed significant association with 6 SOCs; yet, some other cells, such as the 106 cells in Group 1, showed significant association with only 2 SOCs. This implies that (a) there are cells where land price is highly associated with SOCs, and (b) there are cells where land price is least associated with SOCs. Among the cells where SOCs are significantly associated with land price, it is well explained because people would like to live in better living conditions (Choi, Jeung and Park 2021). As for the cells where SOCs are less significantly associated with land price, it can be assumed that the land itself is valuable regardless of SOCs.

As such, this study tried to bring implication for policymakers based on how the selected factors (i.e., level of SOCs such as time and distance, and land usage around SOCs) are associated with the land price. Particularly, a recent plan in Seoul (i.e., 2040 Seoul Development Plan) emphasized the right-for-walking approach, which is a similar concept to the distance to SOCs in this study. The plan has two major goals. The first goal is to help Seoul citizens access the workplace, shopping, and recreation within daily life spaces such as walkable areas. The second goal is to keep the average living quality at a high level across all Seoul areas by achieving the first goal. As the plan emphasized "walkable" accessibility, the result from this study helps to understand how to allocate SOCs across a selected area. Specifically, the understanding of the heterogeneity of residents in a district is recommended to consider when governing entities plan urban management.

The last point is the limitations of the current study. First, the methodological limitation is still existed even if the FMM is strong at mathematical estimation with concurrent performing of clustering and regression. The FMM model may produce different findings by having a different selection of clustering criteria and influential factors. It is the reason why this study employed the hedonic model to select the factors as the theoretical background. Besides the methodological limitation, the empirical approach of the study has some limitation. Specifically, governing entities such as urban planning administrators of the district would intentionally place some SOCs in a certain area where the land price is decreasing. By placing SOCs, the selected areas can be improved and the land price can be kept stable. This means that land price is not solely affected by the effect of SOCs. The SOCs can be affected be the land price as well. However, if the current study includes the reverse association that the land price affects the selection of SOCs, the results may have repeated and recursive questions. Therefore, the current study focused only on the association of the directional effect from SOCs to land price. It was an identification strategy to find the association but brought the study's limitation to exclude other factors' effects. In addition, further investigation on behavioral responses to the land usage by the residents within the cells in Group 1 would be recommended. Furthermore, considering that the hedonic model utilized various internal factors as the influential factors for land price, the next research is recommended to secure a better database with various characteristics. For instance, the price for commuting to the center of a metropolitan area needs to be considered in the next research. As such, by further investigation of different types of cells with spatial heterogeneity, policymakers and administrators may show better performance in urban planning and management.

### 참고문헌

1. Biernacki, C., Celeux, G. and Govaert, G. 1999. An improvement of the NEC criterion for assessing the number of clusters in a mixture model. *Pattern Recognition Letters* 20, no.3: 267-272.
2. Brown, M. E. and Livermore, M. 2019. Identifying individual social capital profiles in low-resource communities: Using cluster analysis to enhance community engagement. *Journal of the Society for Social Work and Research* 10, no.4: 477-500.
3. Buddhavarapu, P., Scott, J. G. and Prozzi, J. A. 2016. Modeling unobserved heterogeneity using finite mixture random parameters for spatially correlated discrete count data. *Transportation Research Part B: Methodological* 91: 492-510.
4. Kim, S. 2021. Soaring home prices stroke anger against Korea's president. Bloomberg Businessweek. July 28. https://www.bloomberg.com/news/articles/2021-07-28/home-prices-soar-in-south-korea-endangering-moon-jae-in-s-reelection (accessed 3 22, 2022).
5. Kim, S., Kim, T. and Suh, K. 2016.Analysis of the implication of accessibility to community facilities for land rrice in rural areas using a hedonic land price model. *Journal of the Korean Society of Rural Planning*. 22, no.1: 93-100.
6. Choi, Y., Jeung, I. and Park, J. 2021. Comparative analysis of spetial impact of living social overhead capital on housing price by residential type. T*ransporting Engineering* 25, 1056-1065. https://doi.org/10.1007/s12205-021-1250-z
7. Cootner, P. H. 1963. Social overhead capital and economic growth. In Rostow, W. W. ed. The economics of take-off into sustained growth. *International Economic Association Series*: 261-184. Palgrave Macmillan. https://doi.org/10.1007/978-1-349-00226-9_15
8. Cottineau, C., Finance, O., Hatna, E., Arcaute, E. and Batty, M. 2019. Defining urban clusters to detect agglomeration economies. *Environment and Planning B: Urban Analytics and City Science* 46, no.9: 1611–1626.
9. Dong, L., Li, R., Zhang, J. and Di, Z. 2016. Population-weighted efficiency in transportation networks. *Scientific Report* 6, article no.26377.
10. Faghih-Imani, A. and Eluru, N. 2017. A finite mixture modeling approach to examine New York City bicycle sharing system (CitiBike) users' destination preferences. *Transportation* 47: 529-553.
11. Fotheringham, A. S. and Pitts, T. C. 1995. Directional variation in distance decay. *Environment and Planning A: Economy and Space* 27, no.5: 715–729.
12. Granovetter, M. 1985. Economic action and social structure: The problem of embeddedness. *American Journal of Sociology* 91, no.3: 481–510.
13. Glumac, B., Herrera-Gomez, M. and Licheron, J. 2019. A hedonic urban land price index. *Land Use Policy* 81: 802-812.
14. Hair, J. F., Sarstedt, M., Ringle, C. M. and Gudergan, S. P. 2018. Advanced Issues in Partial Least Squares Structural Equation Modeling. Sage.

15. Jiao, L. and Liu, Y. 2012. Analyzing the spatial autocorrelation of regional urban datum land price. *Geo-spatial Information Science* 15, no.4: 263-269.
16. Kurihara, K. K. 1970. Social overhead capital and balanced economic growth. *Social and Economic Studies* 19, no.3: 398-405.
17. Lanza, S. T. and Rhoades, B. L. 2013. Latent class analysis: An alternative perspective on subgroup analysis in prevention and treatment. *Prevention Science* 14, no.2: 157–168.
18. Leisch, F. 2004. FlexMix: A general framework for finite mixture models and latent class regression in R. *Journal of Statistical Software* 11, no.8: 1–18.
19. Liu, Q., Deng, M., Shi, Y. and Wang, J. 2012. A density-based spatial clustering algorithm considering both spatial proximity and attribute similarity. *Computers & Geosciences* 46: 296-309.
20. McLachlan, G. J. 2009. Model-Based Clustering. In Brown, S. D., R. Thauler and B. Walczak. eds. *Comprehensive Chemometrics*: 655-681. Elsevier.
21. Maddison, J. D. 2000. A hedonic analysis of agricultural land prices in England and Wales. *European Review of Agricultural Economics* 27, no.4: 519-532.
22. Ministry of Land, Infrastructure and Transport. 2019. Ministry of Land, Infrastructure a Guideline to Analyze the Report. A Report of Supplying Basic Infrastructure and Transport. Retrieved from https://www.codil.or.kr/filebank/original/RK/OTKCRK190154/OTKCRK190154.pdf?stream=T (accessed 4 26, 2022).
23. Naughton, L. 2013. Geographical narratives of social capital: Telling different stories about the socio-economy with context, space, place, power and agency. *Progress in Human Geography* 38, no.1: 3-21.
24. Nest, G., Passos, V. L., Candel, M. J. J. M. and Breukelen, G. J. P. (2020). An overview of mixture modelling for latent evolutions in longitudinal data: Modeling approaches, fit statistics and software. *Advances in Life Course Research* 43: 100323. https://doi.org/10.1016/j.alcr.2019.100323
25. Nocco, A. 2009. Preference heterogeneity and economic geography. *Journal of Regional Science* 49, no.1: 33-56.
26. Oller, R., Martori, J. C. and Madariaga, R. 2016. Monocentricity and directional heterogeneity: A conditional parametric approach. *Geographical Analysis* 49, no.3: 343-361.
27. Ottaviano, G. I. P. 2011. 'New' new economic geography: Firm heterogeneity and agglomeration economies. *Journal of Economic Geography* 11, no.2: 231-240.
28. Oxford Reference. n.d. Social Overhead Capital. In Oxfordreference.com Dictionary. Retrieved from https://www.oxfordreference.com/view/10.1093/oi/authority.20110803100515344 (accessed 5 25, 2022).
29. Parr, J. B. 2007. Spatial definitions of the city: Four perspectives. *Urban Studies* 44, no.2: 381–392.
30. Peel, D. and McLachlan, G. J. 2000. Robust mixture modelling using the t distribution. *Statistics and Computing* 10, no.4: 339–348.
31. Rosen, S. 1974. Hedonic prices and implicit markets: Product differentiation in pure competition. *Journal of Political Economy* 82, no.1: 34-55.
32. Sampson, R. J. and Graif, C. 2009. Neighborhood social capital as differential social organization: Resident and leadership dimensions. *American Behavioral Scientist* 51, no.11: 1579-1605.
33. Sarstedt, M. 2008. A review of recent approaches for capturing heterogeneity in partial least squares path modelling. *Journal of Modeling in Management* 3, no.2: 140-161.
34. Sheppard, E. 2011. Geography, nature, and the question of development. *Dialogues in Human Geography* 1, no.1: 46–75.
35. Simon, S. 2009. A modern approach to regression with R. *Springer Texts in Statistics*. Springer.
36. StataCorp. 2021. Stata Finite Mixture Models Reference Manual. Release 17. A Stata Press Publication.

37. Suh, K. 2005. Analysis of determinant factors of land price in rural area using a hedonic land price model and spatial econometric models. *Journal of the Korean Society of Rural Planning* 11, no.3: 11-17.
38. Seong, E. Y., Lim, Y. J. and Lim, H. S. 2013. Accessibility and Availability of Neighborhood Facilities in Old Residential Area. Architecture and Urban Research Institute. Retrieved from www.auric.or.kr/User/Rdoc/DocRdoc.aspx?returnVal=RD_R&dn=321291#.X9HqG6Gg-UI (accessed 3 21, 2022).
39. Tabuchi, T. and Thisse, J. 2002. Taste heterogeneity, labor mobility and economic geography. *Journal of Development Economics* 69, no.1: 155-177.
40. Vermunt, J. K. and Magidson, J. 2002. Latent class cluster analysis. *Applied Latent Class Analysis* 11: 89–106.
41. Wedel, M. and Kamakura, W. A. 2000. Market Segmentation: Conceptual and Methodological Foundations. Kluwer Academic Publishers.

## 요약

주제어: 공간적 이질성, 유한혼합모델링, 사회간접자본, 토지가격, 공간분석

SOC(Social Overhead Capitals, 사회간접자본)가 지역사회의 토지 가치와 어떻게 연관되어 있는지 이해하는 것은 효과적인 도시 계획을 추진함에 있어 중요한 요소 중 하나이다. 하지만 지역 내에서 SOC가 여러 가지 목적으로 사용되는 부분이 있는데 이것을 공간적 이질성이라 한다. 지가를 파악하기 위해서는 공간적 이질성 문제를 고려해야 한다. 공간적 이질성이 구역 내에 있는 경우 공간 클러스터링 방식을 통하여 지가 정책을 관리할 수 있다. 본 연구에서는 (a) 최적의 클러스터 수 및 (b) SOC 간의 연관성을 찾기 위해 FMM(Finite Mixture Modeling, 유한혼합모델링)을 사용하여 특정 지역의 SOC, 사회인구학적 특성 및 공간 정보를 포함한 공간 속성을 분석하였다. FMM 분석 방법은 클러스터와 속성의 계수를 동시에 분석하는데 사용되는 빅데이터 분석방법으로, 사회인구통계학적 특징 및 토지 가격 지표 등 다양한 자료를 분석하는데 해당 분석방법을 활용하였다. 분석 방법인 FMM은 클러스터를 구성하면서 회귀계수를 동시에 추정하는 방식이다. FMM 방법을 사용한 결과는 4개의 클러스터가 하나의 연구 대상지에 존재하고, 4개의 클러스터는 SOC, 인구학적 특성 및 지가 간에 서로 다른 연관성을 가지고 있음을 보여주었다. 정책 입안자와 관리 행정기관은 토지 가격 정책을 만들기 위해 다양한 도시 지표 등의 정보 발굴 및 분석이 필요하다. 본 연구에서는 SOC와의 근접성을 고려하는 것이 지가에 중요한 요인임을 확인하였으며, 향후 SOC와 관련된 계획 및 정책의 개선 방향을 제시하였다.

# Appendix •••••

**Appendix _** Features Summary of Four Groups

| Category | | Group 1<br>(n = 389)<br>Mean (S.D.) | Group 2<br>(n = 265)<br>Mean (S.D.) | Group 3<br>(n = 144)<br>Mean (S.D.) | Group 4<br>(n = 106)<br>Mean (S.D.) |
|---|---|---|---|---|---|
| Land price (won) | | Low | Low | High | High |
| Population | | Average | Average | Average | Low |
| Female rate | | Average | Average | Average | Average |
| Rate of public land | | Low | Low | High | High |
| Land Utilities | | Residential | Residential | Complex | Complex |
| Distance to SOCs | | | | | |
| | Kindergarten | Short | Short | Far long | Long |
| | Elementary school | Long | Short | Short | Short |
| | Public library | Short | Short | Far long | Far long |
| | Daycare | Short | Short | Long | Far long |
| | Senior community | Long | Short | Very short | Long |
| | Senior education | Short | Short | Long | Far long |
| | Health facilities | Short | Short | Long | Far long |
| | Neighborhood park | Short | Short | Far long | Long |
| | Public park | Short | Long | Long | Short |